\documentclass[12pt]{iopart}

\usepackage{graphicx}
\begin{document}

\title[]{Synthesis and physical properties of perovskite Sm$_{1-x}$Sr$_x$NiO$_3$ (\emph{x} = 0, 0.2) and infinite-layer Sm$_{0.8}$Sr$_{0.2}$NiO$_2$ nickelates}

\author{Chengping He, Xue Ming, Qing Li, Xiyu Zhu$^{*}$, Jin Si and Hai-Hu Wen$^{*}$}

\address{Center for Superconducting Physics and Materials, National Laboratory of Solid State Microstructures and Department of Physics, Collaborative Innovation Center of Advanced Microstructures, Nanjing University, Nanjing 210093, China}
\ead{zhuxiyu@nju.edu.cn and hhwen@nju.edu.cn}
\vspace{10pt}
\begin{indented}
\item[]January 2021
\end{indented}

\begin{abstract}
Recently, superconductivity at about 9-15 K was discovered in Nd$_{1-x}$Sr$_x$NiO$_2$ (Nd-112, \emph{x} $\approx$ 0.125-0.25) infinite-layer thin films, which has stimulated enormous interests in related rare-earth nickelates. Usually, the first step to synthesize this 112 phase is to fabricate the \emph{R}NiO$_3$ (\emph{R}-113, \emph{R}: rare-earth element) phase, however, it was reported that the 113 phase is very difficult to be synthesized successfully due to the formation of unusual Ni$^{3+}$ oxidation state. And the difficulty of preparation is enhanced as the ionic radius of rare-earth element decreases. In this work, we report the synthesis and investigation on multiple physical properties of polycrystalline perovskites Sm$_{1-x}$Sr$_x$NiO$_3$ (\emph{x} = 0, 0.2) in which the ionic radius of Sm$^{3+}$ is smaller than that of Pr$^{3+}$ and Nd$^{3+}$ in related superconducting thin films. The structural and compositional analyses conducted by X-ray diffraction and energy dispersive X-ray spectrum reveal that the samples mainly contain the perovskite phase of Sm$_{1-x}$Sr$_x$NiO$_3$ with small amount of NiO impurities. Magnetization and resistivity measurements indicate that the parent phase SmNiO$_3$ undergoes a paramagnetic-antiferromagnetic transition at about 224 K on a global insulating background. In contrast, the Sr-doped sample Sm$_{0.8}$Sr$_{0.2}$NiO$_3$ shows a metallic behavior from 300 K down to about 12 K, while below 12 K the resistivity exhibits a slight logarithmic increase. Meanwhile, from the magnetization curves, we can see that a possible spin-glass state occurs below 12 K in Sm$_{0.8}$Sr$_{0.2}$NiO$_3$. Using a soft chemical reduction method, we also obtain the infinite-layer phase Sm$_{0.8}$Sr$_{0.2}$NiO$_2$ with square NiO$_2$ planes. The compound shows an insulating behavior which can be described by the three-dimensional variable-range-hopping model. And superconductivity is still absent in the polycrystalline Sm$_{0.8}$Sr$_{0.2}$NiO$_2$.
\end{abstract}

%
%
%
%
%

\section{Introduction}

The discovery of high critical temperature superconductivity (HTS) in cuprates in 1986 \cite{LaBaCuO} created a new era for research of superconductivity. However, after great efforts by many scientists, there are still a lot of puzzlings about the superconducting mechanism of cuprates \cite{review}. One of the effective ways to resolve these puzzlings is to explore other new superconductors which are isostructural and isoelectronic to the cuprates \cite{Ni438, LaNiO3/LaMO3}. Considering the proximity of nickel and copper in the periodic table, it is natural to raise the question that whether the cuprate-like electronic structure or even superconductivity could be realized in nickelates. If so, this would help to understand the nature of HTS in cuprates. Early theoretical studies proposed that layered nickelates with square planar NiO$_2$ sheets and the Ni$^{1+}$ oxidation state (supposed to give an S = 1/2 magnetic moment) may be analogous to the parent phase of cuprates. By doping holes into the system, superconductivity like that in cuprate superconductors may be realized in nickelates \cite{TW1, TW2}. And this has come true experimentally in Nd$_{1-x}$Sr$_x$NiO$_2$ (0.125 $\textless$ \emph{x} $\textless$ 0.25) \cite{NdSrNiO, phase diagram, phase diagram2} and Pr$_{0.8}$Sr$_{0.2}$NiO$_2$ \cite{PrSrNiO} thin films deposited on SrTiO$_3$ substrates with transition temperatures at about 9-15 K and 7-12 K, respectively. Since the superconductivity in the infinite-layer nickelates was discovered in thin films, it is of great interest to know whether it can be realized in bulk form. However, the recent transport measurements on bulk Nd$_{1-x}$Sr$_x$NiO$_2$ samples show insulating behaviors at both ambient and high pressures \cite{NdSrNiO bulk, NdSrNiO bulk2}. Thus the key question is to know what makes the results in thin films and in bulk samples so contradictory. Since the Nd$_{1-x}$Sr$_x$NiO$_2$ and Pr$_{0.8}$Sr$_{0.2}$NiO$_2$ thin films were grown on SrTiO$_3$ substrates, one plausible explanation is that the substrates may play a crucial role in the emergence of superconductivity \cite{absence, substrate effect, Progress}. As mentioned by Li et al. \cite{NdSrNiO}, the films experience compressive strain on the SrTiO$_3$ substrates. And the smaller lattice constant may be more conducive to superconductivity \cite{smaller lattice constant}. Following this idea, it may be a good way to simulate the strain effect of substrates on the films by replacing Nd$^{3+}$ or Pr$^{3+}$ with smaller rare-earth cation Sm$^{3+}$.

Looking back to literatures, it is found that the \emph{R}NiO$_3$ perovskites have been intensively studied \cite{La1, Nd, Sm and Pr, Gd and Dy, R1, Y, R2} due to their metal-insulator (MI) transitions and paramagnetic-antiferromagnetic transitions. Except for LaNiO$_3$ which has a different crystal structure (rhombohedral) from other \emph{R}NiO$_3$ and is a paramagnetic metal in the whole measured temperature range \cite{La2}, \emph{R}NiO$_3$ perovskites are paramagnetic metals at high temperatures, while they are in an insulating and antiferromagnetic state at low temperatures \cite{R3}. For \emph{R} = Pr and Nd, the MI transition temperature (\emph{T}$_{{\rm MI}}$) is the same as the paramagnetic-antiferromagnetic transition temperature (\emph{T}$_{\rm N}$) \cite{Nd and Pr}. However, for \emph{R} = Sm or smaller rare-earth ions, \emph{T}$_{{\rm MI}}$ is usually higher than \emph{T}$_{\rm N}$. In the region between \emph{T}$_{{\rm MI}}$ and \emph{T}$_{\rm N}$, \emph{R}NiO$_3$ are paramagnetic insulators. The value of \emph{T}$_{{\rm MI}}$ is strongly dependent on the ionic radius of the rare-earth ions \cite{R4}. As the ionic radius decreases, the tolerance factor for stabilizing a regular perovskite structure becomes smaller. The decreased tolerance factor further lead to the reduction of Ni-O-Ni bond angle. Thus the bandwidth which is composed of Ni-3\emph{d} and O-2\emph{p} orbitals \cite{Bi} is also reduced. In this case, the physical properties of \emph{R}NiO$_3$ can be tuned by a variety of methods such as the displacement of rare-earth elements, chemical doping, pressure, temperature, and so on. For SmNiO$_3$, it has been reported that \emph{T}$_{{\rm MI}}$ and \emph{T}$_{\rm N}$ are 403 K and 225 K, respectively \cite{Sm and Pr}. And crystal symmetry changes from the high-temperature orthorhombic structure (space group: Pbnm) to the low-temperature monoclinic structure (space group: P2$_1$/n) \cite{Progress, Sm1, Sm add}. High pressure studies have shown that the pressure is beneficial to the metallic state and drives the MI transition temperature to a lower value \cite{Sm2, Sm3}. Furthermore, due to the difficulty in the fabrication of bulk samples, previous researches mainly focused on thin films and heterostructures \cite{Progress, R3}. To our knowledge, there is no study on the doping effect of bulk SmNiO$_3$. Therefore, it is interesting to know whether it will show some novel physical properties upon doping in bulk samples.

Motivated by the reasonings above, in this paper, we report the synthesis and investigation on multiple physical properties of Sm$_{1-x}$Sr$_x$NiO$_3$ (\emph{x} = 0, 0.2) and Sm$_{0.8}$Sr$_{0.2}$NiO$_2$ polycrystalline samples. The powder X-ray diffraction (XRD) measurements have confirmed the perovskite structures for Sm$_{1-x}$Sr$_x$NiO$_3$ (\emph{x} = 0, 0.2) and infinite-layer structure for Sm$_{0.8}$Sr$_{0.2}$NiO$_2$. The compositional analyses show that the Ni occupancy is slightly lower than the nominal value in all three samples and there is a certain strontium distribution in Sm$_{0.8}$Sr$_{0.2}$NiO$_3$ and Sm$_{0.8}$Sr$_{0.2}$NiO$_2$. For SmNiO$_3$, the paramagnetic-antiferromagnetic transition appears at about 224 K on an insulating background. Upon Sr doping, Sm$_{0.8}$Sr$_{0.2}$NiO$_3$ exhibits a metallic behavior accompanied with a resistivity upturn at low temperatures, which seems to be fitted fairly well by the picture concerning magnetic Kondo scattering. The magnetization curves of Sm$_{0.8}$Sr$_{0.2}$NiO$_3$ show paramagnetic behaviors with a possible spin-glass state below 12 K. After doing soft chemical reduction from Sm$_{0.8}$Sr$_{0.2}$NiO$_3$, we successfully synthesize Sm$_{0.8}$Sr$_{0.2}$NiO$_2$ samples. The Sm$_{0.8}$Sr$_{0.2}$NiO$_2$ shows an insulating behavior which can be well fitted with three-dimensional (3D) variable-range-hopping (VRH) model (ln$\rho$ $\propto$ \emph{T}$^{-1/4}$) in the whole measured temperature range. There is no sign of superconductivity above 2 K in the polycrystalline Sm$_{0.8}$Sr$_{0.2}$NiO$_2$.

\section{Experimental methods}

Polycrystalline samples of Sm$_{1-x}$Sr$_x$NiO$_3$ (\emph{x} = 0, 0.2) were successfully synthesized under high pressure and high temperature by solid-state reaction method. First, the precursors Sm$_{2-2x}$Sr$_{2x}$NiO$_4$ (\emph{x} = 0, 0.2) were prepared by calcining stoichiometric amounts of Sm$_2$O$_3$ (99.9$\%$, Alfa Aesar), NiO (99.0$\%$, Alfa Aesar) and SrO (99.9$\%$, Strem Chemicals) at 1200 $^{\circ}$C for 24 h. Then the precursors Sm$_{2-2x}$Sr$_{2x}$NiO$_4$, NiO and KClO$_4$ (99.0$\%$, Alfa Aesar) were weighed in a molar ratio of 1:1:2, mixed and ground thoroughly in an agate mortar. Subsequently, the mixture was pressed into pellets and sealed in a gold capsule. These procedures were carried out in a glove box filled with purified Ar gas (H$_2$O, O$_2$ \textless 0.1 ppm). The gold capsule with the mixed compounds was then placed in a BN capsule and heated up to 1000 $^{\circ}$C and kept at this temperature for 2 h under the pressure of 2 GPa in a piston-cylinder high pressure apparatus (LPC 250-300/50, Max Voggenreiter). Then the samples were rapidly cooled to room temperature in 1 min followed by the release of pressure. The obtained samples were washed with distilled water to dissolve KCl and unreacted KClO$_4$, and then dried.

The sample Sm$_{0.8}$Sr$_{0.2}$NiO$_2$ was obtained by reacting Sm$_{0.8}$Sr$_{0.2}$NiO$_3$ with CaH$_2$ (98.5$\%$, Aladdin) through a soft chemical reduction method. Appropriate amounts of Sm$_{0.8}$Sr$_{0.2}$NiO$_3$ and 0.1 g of CaH$_2$ powder were pressed into pellets respectively and then sealed in an evacuated quartz tube. Note that the Sm$_{0.8}$Sr$_{0.2}$NiO$_3$ and CaH$_2$ pellets were placed at different positions in the quartz tube to avoid  direct contact. The quartz tube was heated to 340 $^{\circ}$C at a rate of 10 $^{\circ}$C/min and sustained at this temperature for 10 h. We have also tried to reduce SmNiO$_3$ with CaH$_2$ in order to obtain SmNiO$_2$, but failed probably because of the instability of Ni$^{1+}$ oxidation state \cite{NdSrNiO bulk2}.

The phase identification of the prepared samples was carried out by powder XRD (Bruker D8 Advance) with Cu-K$\alpha$ radiation and the TOPAS 4.2 software \cite{TOPAS} was used to refine the crystal structures by Rietveld analysis \cite{Rietveld}. Micrographs of the samples were taken by a scanning electron microscopy (SEM) (Phenom ProX, Phenom World) and the chemical composition analyses were done with an energy dispersive X-ray spectrum (EDS) which is an option of the SEM. The dc magnetization was collected on a vibrating sample magnetometer (PPMS-VSM 9T, Quantum Design). The electrical resistivities of the samples were measured by the standard four-probe method using the physical property measurement system (PPMS 16T, Quantum Design).

\section{Results and discussion}
\subsection{Sm$_{1-x}$Sr$_x$NiO$_3$ (x = 0, 0.2)}
\subsubsection{Sample characterization}

\begin{figure}
  \centering
  \includegraphics[width=3in]{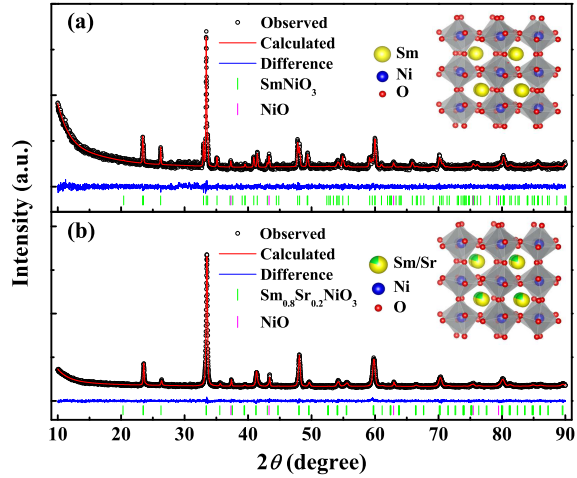}
  \caption{Powder X-ray diffraction patterns (black circles) and Rietveld refinement fitting curves (red lines) of (a) SmNiO$_3$ and (b) Sm$_{0.8}$Sr$_{0.2}$NiO$_3$. The insets of (a) and (b) show the schematic crystal structures of SmNiO$_3$ and Sm$_{0.8}$Sr$_{0.2}$NiO$_3$, respectively.}\label{fig1}
\end{figure}

\begin{table}
\caption{\label{tbl}Crystallographic data of SmNiO$_3$ and Sm$_{0.8}$Sr$_{0.2}$NiO$_3$ at room temperature.}
\begin{center}
\item[]\begin{tabular}{@{}llll}
\br
fomula & SmNiO$_3$ & Sm$_{0.8}$Sr$_{0.2}$NiO$_3$ \\
\mr
space group & $Pbnm$ & $Pbnm$\\
$a$ ($\mathring{A}$) & 5.3241(6) & 5.3641(8) \\
$b$ ($\mathring{A}$) & 5.4282(2) & 5.3564(8) \\
$c$ ($\mathring{A}$) & 7.5594(7) & 7.5675(8) \\
$V$ ($\mathring{A}^3$) & 218.4743(9) & 217.4405(1) \\
$\rho$ (g/cm$^3$) & 7.815 & 7.469 \\
$R_{wp}$ (\%) & 4.80 & 3.24 \\
$R_{p}$ (\%) & 3.77 & 2.55 \\
$GOF$ & 1.10 & 1.11 \\
\br
\end{tabular}
\end{center}
\end{table}

Figure~\ref{fig1} shows the powder XRD patterns of (a) SmNiO$_3$ and (b) Sm$_{0.8}$Sr$_{0.2}$NiO$_3$ polycrystalline samples with Rietveld refinements. Except for several minor peaks of NiO impurities, all other peaks belong to the perovskite SmNiO$_3$ and Sm$_{0.8}$Sr$_{0.2}$NiO$_3$ phases which can be well indexed with an orthorhombic unit cell (space group: Pbnm). Their schematic crystal structures have been illustrated in the insets of Fig.~\ref{fig1}(a) and Fig.~\ref{fig1}(b), where the yellow/green, blue and red spheres represent the Sm/Sr, Ni and O atoms, respectively. The perovskite structure is distorted because of the relative small ionic radius of Sm$^{3+}$ cation, and the NiO$_6$ octahedra are tilted for optimizing the Sm-O distances. The corner-sharing octahedra are arranged along three directions of the crystal and  Sm$^{3+}$ cations occupy the space between these NiO$_6$ octahedra. The cell parameters obtained from Rietveld refinement profiles are \emph{a} = 5.3241(6) $\mathring{A}$, \emph{b} = 5.4282(2) $\mathring{A}$, \emph{c} = 7.5594(7) $\mathring{A}$ for SmNiO$_3$ and \emph{a} = 5.3641(8) $\mathring{A}$, \emph{b} = 5.3564(8) $\mathring{A}$, \emph{c} = 7.5675(8) $\mathring{A}$ for Sm$_{0.8}$Sr$_{0.2}$NiO$_3$, which are consistent with previous studies \cite{R1, Sm abc1, Sm abc2}. It can be seen that the lattice constant \emph{c} expands slightly after Sr doping, which is reasonable since the ionic radius of Sr$^{2+}$ (118 pm) is larger than that of Sm$^{3+}$ (95.8 pm) \cite{Sr and Sm ionic radius}. The crystallographic parameters and the reliability factors have been listed in Table~\ref{tbl}. The small values of R$_{wp}$ and R$_p$ indicate that the fittings are considerably good.

\begin{figure}
  \centering
  \includegraphics[width=4in]{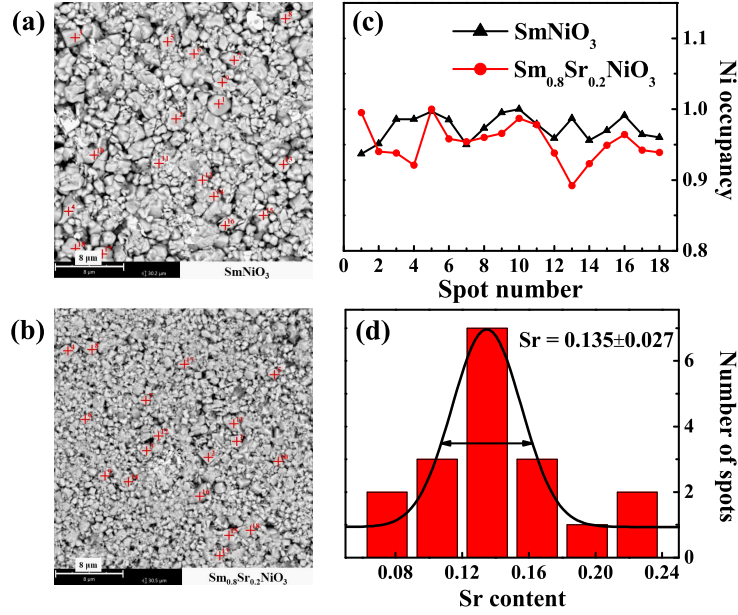}
  \caption{SEM images of (a) SmNiO$_3$ and (b) Sm$_{0.8}$Sr$_{0.2}$NiO$_3$. (c) Ni occupancy in SmNiO$_3$ and Sm$_{0.8}$Sr$_{0.2}$NiO$_3$. (d) Compositional distribution of Sr content with Gaussian fitting for Sm$_{0.8}$Sr$_{0.2}$NiO$_3$.}\label{fig2}
\end{figure}

Figure~\ref{fig2} presents the SEM images of (a) SmNiO$_3$ and (b) Sm$_{0.8}$Sr$_{0.2}$NiO$_3$ samples. It can be seen clearly that both samples contain two types of grains, which are the major part of the grains with argenteous color and a small amount of dark gray ones. Then we perform EDS analyses to get a deeper understanding of the element composition. We measure 18 spots for both samples randomly, which have been marked by the red crosses in the two images. Note that we prepared the samples in stoichiometry. However, by analyzing the Sr content and Ni occupancy normalized by the concentration of Sm for SmNiO$_3$ and (Sm+Sr) for Sm$_{0.8}$Sr$_{0.2}$NiO$_3$, we find that both of them deviate slightly from the nominal values. Namely, the argenteous grains correspond to the composition of SmNi$_{1-\delta}$O$_3$ for SmNiO$_3$ and Sm$_{1-y}$Sr$_{y}$Ni$_{1-\delta}$O$_3$ for Sm$_{0.8}$Sr$_{0.2}$NiO$_3$, respectively. And the small amount of dark gray grains are NiO, which is consistent with the XRD results. Fig.~\ref{fig2}(c) presents the Ni occupancy of different spots for SmNiO$_3$ and Sm$_{0.8}$Sr$_{0.2}$NiO$_3$. The Ni occupancy changes slightly and the mean values are about 0.97(4) for SmNiO$_3$ and 0.95(3) for Sm$_{0.8}$Sr$_{0.2}$NiO$_3$. Both values are slightly lower than the nominal one. Note that R. Jaramillo et al. \cite{Sm add2} did EDS measurements on SmNiO$_3$ thin films and found the ratio of Sm to Ni is around 1.1 with an error of 0.1 but without the indication of Ni deficiency. Since the accuracy of EDS measurement depends on the calibration of the system, sample morphology and the actual location of the energy spectrum from the calculated elements, there are experimental errors in EDS analysis. Thus, whether there is Ni deficiency in the samples requires further experiments. Fig.~\ref{fig2}(d) shows the compositional distribution of Sr content with Gaussian fitting for Sm$_{0.8}$Sr$_{0.2}$NiO$_3$. The Sr content is randomly distributed in Sm$_{0.8}$Sr$_{0.2}$NiO$_3$ and the average value obtained from the fitting is 0.135$\pm$0.027, which is close to the nominal value and implies the successful doping of Sr. In addition, the full width at half maximum (FWHM) of the XRD peak at around 26$^{\circ}$ for Sm$_{0.8}$Sr$_{0.2}$NiO$_3$ is 0.24, which is larger than the value of 0.15 for SmNiO$_3$. The broader peak indicates smaller grains in Sm$_{0.8}$Sr$_{0.2}$NiO$_3$ or the Sr concentration has a distribution in the grains. Since the spot number in Fig.~\ref{fig2}(c) is sorted by the increase of Sr content, there is no apparent correlation between Sr content and Ni occupancy among different spots.

\subsubsection{Magnetic and electrical transport properties}

\begin{figure}
  \centering
  \includegraphics[width=4in]{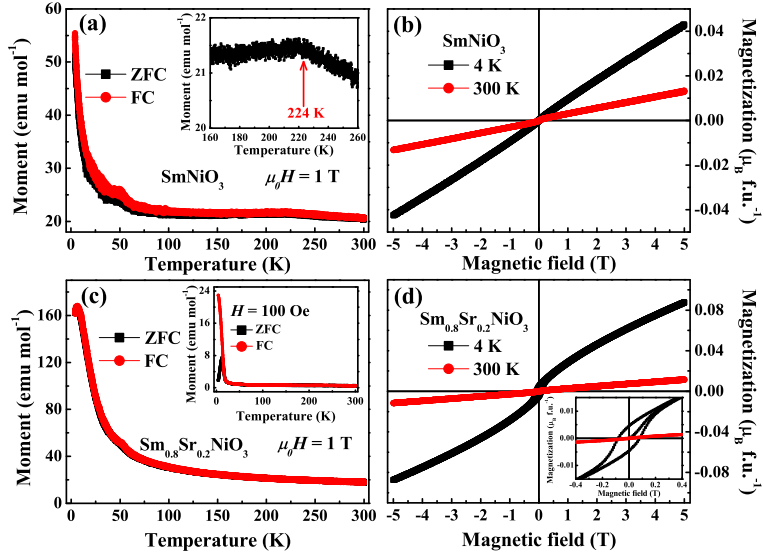}
  \caption{Temperature dependences of magnetic moment measured in both ZFC and FC modes at 1 T for (a) SmNiO$_3$ and (c) Sm$_{0.8}$Sr$_{0.2}$NiO$_3$. The insets of (a) and (c) show the enlarged view around the paramagnetic-antiferromagnetic transition for SmNiO$_3$ and  temperature dependences of magnetic moment in ZFC and FC modes at 100 Oe for Sm$_{0.8}$Sr$_{0.2}$NiO$_3$, respectively. Magnetization hysteresis loops measured at 4 K and 300 K for (b) SmNiO$_3$ and (d) Sm$_{0.8}$Sr$_{0.2}$NiO$_3$. The inset of (d) shows the zoom-in view of the loops at low fields for Sm$_{0.8}$Sr$_{0.2}$NiO$_3$.} \label{fig3}
\end{figure}

Figure~\ref{fig3}(a) displays the temperature dependences of magnetic moment at the field of 1 T in zero-field-cooling (ZFC) and field-cooling (FC) modes for SmNiO$_3$. As we can see, the magnetic moment increases smoothly with decreasing temperature from 300 K to about 224 K and a small kink is observed at 224 K (see the enlarged ZFC curve in the inset of Fig.~\ref{fig3}(a)), which is consistent with previous reports and has been attributed to the paramagnetic-antiferromagnetic transition \cite{Sm and Pr, Sm1}. An abnormal hump-like behavior is observed at about 50 K, which is attributed to the antiferromagnetic transition of frozen oxygen. We have also measured the magnetization hysteresis at 4 K and 300 K for SmNiO$_3$ and present the results in Fig.~\ref{fig3}(b). Both curves are roughly linear and the magnetization is unsaturated up to the field of 5 T, which rule out the possibility of ferromagnetic component. After Sr doping, the magnetic moment of Sm$_{0.8}$Sr$_{0.2}$NiO$_3$ increases with temperature cooling down, while a drop appears below 7 K, which are shown in Fig.~\ref{fig3}(c). To illustrate the small drop at low temperatures, we measure the temperature dependences of magnetic moment in ZFC and FC modes under the field of 100 Oe, and present the results in the inset of Fig.~\ref{fig3}(c). A clear magnetic irreversibility occurs below 12 K between ZFC and FC curves. Besides, in Fig.~\ref{fig3}(d), we perform the magnetization hysteresis loops measured at 4 K and 300 K for Sm$_{0.8}$Sr$_{0.2}$NiO$_3$. The linear \emph{M}(\emph{H}) relation at 300 K demonstrates a paramagnetic like behavior. However, an obvious deviation from linearity is observed from the curve of Sm$_{0.8}$Sr$_{0.2}$NiO$_3$ at 4 K. The small magnetic hysteresis demonstrates a weak ferromagnetic behavior. The weak ferromagnetic behavior at low temperatures seen in \emph{M}-\emph{H} curves has been also observed in BiNiO$_3$ \cite{Bi} and PbNiO$_3$ \cite{Pb}, and has been attributed to the canted spins or the magnetic impurity phases. Combined with \emph{M}-\emph{T} and \emph{M}-\emph{H} results in Sm$_{0.8}$Sr$_{0.2}$NiO$_3$, the low-temperature magnetization behaviors observed here are more likely related to a spin-glass state. Similar behaviors have been found in Ca-doped Lu$_{1-x}$Ca$_x$MnO$_3$ single crystals \cite{SG1}, Mn-doped CaNiGeH \cite{SG2} and KCr$_3$As$_3$ \cite{SG3}. But we can not exclude the possibility that the behaviors in our samples are caused by impurities or the canted spins. Further study is needed in order to clarify the origin of the low-temperature magnetization behaviors in Sm$_{0.8}$Sr$_{0.2}$NiO$_3$.

\begin{figure}
  \centering
  \includegraphics[width=3in]{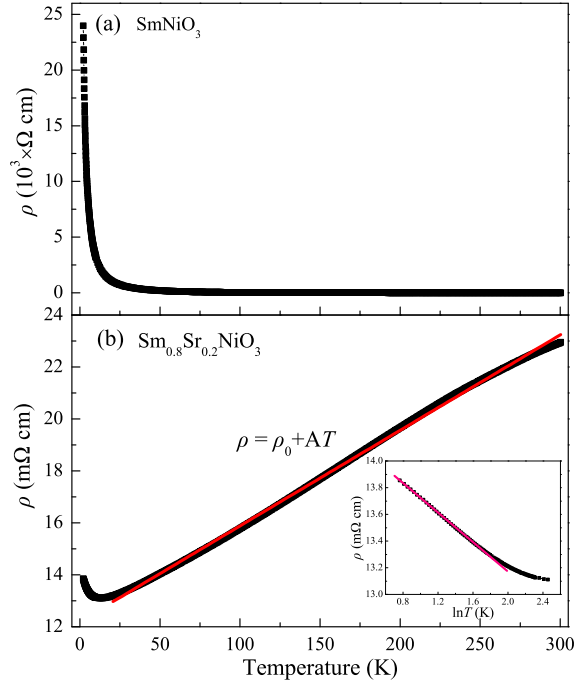}
\caption{Temperature dependences of resistivity for (a) SmNiO$_3$ and (b) Sm$_{0.8}$Sr$_{0.2}$NiO$_3$. The inset of (b) shows the logarithmic temperature dependence of resistivity at low temperature for Sm$_{0.8}$Sr$_{0.2}$NiO$_3$.} \label{fig4}
\end{figure}

Figure~\ref{fig4} shows the comparison of the temperature dependences of resistivity for (a) SmNiO$_3$ and (b) Sm$_{0.8}$Sr$_{0.2}$NiO$_3$. SmNiO$_3$ exhibits an insulating behavior in the whole measured temperature range from 2 K to 300 K, which is consistent with previous reports \cite{R3, Sm and Pr}. In order to better understand the transport behavior of SmNiO$_3$, we have tried to fit the data with several models, such as the band-gap model (ln$\rho$ $\propto$ 1/\emph{T}), small-polaron-hopping (SPH) model (ln($\rho/T) \propto -1/T$) and 3D VRH model (ln$\rho \propto T^{-1/4}$) \cite{LH}, but all failed. We should note that the oxygen nonstoichiometry usually occurs in \emph{R}NiO$_3$ nickelates due to the instability of Ni$^{3+}$ oxidation state \cite{transport4}. And the oxygen deficiency in the sample plays an important role in transport properties, just like the cases in cuprates and other nickelates \cite{transport1, transport2, transport3, transport5}. Furthermore, since the pellets used for resistivity measurements were pressed by powder, the grain boundary is another nonnegligible factor which can affect the transport properties \cite{grain}. These factors may lead to the exotic scattering in SmNiO$_3$ polycrystals which can not be understood by present models and is different from that in thin films, in which the insulating behavior can be fitted by the VRH model \cite{Sm add3}. In contrast, Sm$_{0.8}$Sr$_{0.2}$NiO$_3$ shows a metallic behavior above 12 K with a resistivity upturn at lower temperatures. As we can see in Fig.~\ref{fig4}(b), the $\rho$ -\emph{T} curve is roughly linear between 12 K and 300 K, so we attempt to fit the data using the formula $\rho$(\emph{T}) = $\rho$$_0$ + A\emph{T} and the fitting result has been depicted by the red line. The fitting result yields $\rho$$_0$ = 12.21 m$\Omega$ cm and A = 3.68$\times$10$^{-2}$ m$\Omega$ cm K$^{-1}$. The metallic behavior observed here may be caused by hole doping through Sr substitution, and has also been observed in bulk Nd$_{0.8}$Sr$_{0.2}$NiO$_3$ \cite{NdSrNiO bulk2}. The low-temperature resistivity upturn can be fitted well by the logarithmic function, namely $\rho$ $\propto$ ln\emph{T} (see the inset in Fig.~\ref{fig4}(b)), which can be attributed to the magnetic Kondo scattering. In this scenario, partial localized Ni-3d electrons carry magnetic moments which serve as the Kondo-scattering centers and thus result in the scattering of conducting electrons \cite{Kondo1, Kondo2, Kondo3}. In addition, the grain boundary may also result in the low-temperature resistivity upturn, since the pellets used for resistivity measurements were pressed by powder.

\subsection{Sm$_{0.8}$Sr$_{0.2}$NiO$_2$}
\subsubsection{Sample characterization}

\begin{figure}
  \centering
  \includegraphics[width=4in]{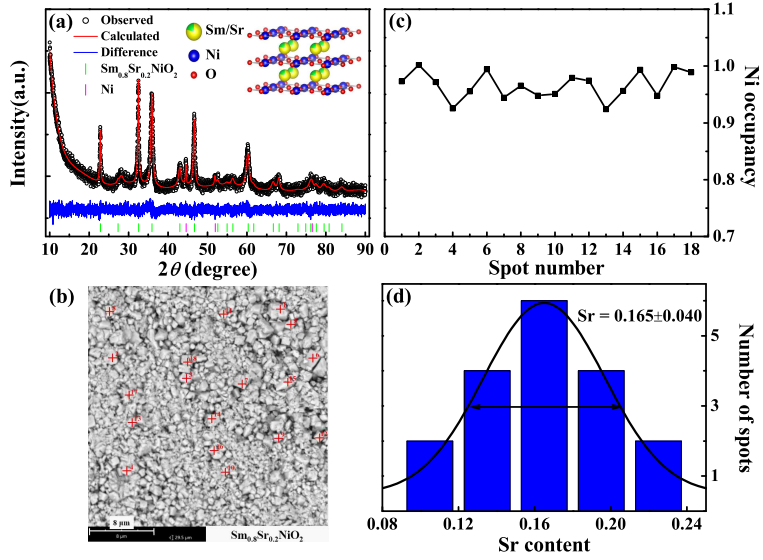}
  \caption{Powder X-ray diffraction pattern (black circles) and Rietveld refinement fitting curve (red line) for Sm$_{0.8}$Sr$_{0.2}$NiO$_2$. The inset shows the crystal structure of Sm$_{0.8}$Sr$_{0.2}$NiO$_2$. (b) SEM image of Sm$_{0.8}$Sr$_{0.2}$NiO$_2$. (c) Ni occupancy and (d) compositional distribution of Sr concentration with Gaussian fitting for Sm$_{0.8}$Sr$_{0.2}$NiO$_2$.} \label{fig5}
\end{figure}

Using a soft chemical reduction method, we successfully obtain the Sm$_{0.8}$Sr$_{0.2}$NiO$_2$ polycrystalline samples from the prepared perovskite Sm$_{0.8}$Sr$_{0.2}$NiO$_3$. Figure~\ref{fig5}(a) displays the powder XRD pattern of Sm$_{0.8}$Sr$_{0.2}$NiO$_2$ sample with Rietveld refinement. And the inset presents the schematic crystal structure. Except for 2-3 minor peaks of Ni impurities, all other peaks correspond to the infinite-layer Sm$_{0.8}$Sr$_{0.2}$NiO$_2$ phase which can be well indexed with a tetragonal unit cell (space group: P4/mmm). Analogous to the case in our reported Nd$_{0.8}$Sr$_{0.2}$NiO$_2$ samples \cite{NdSrNiO bulk}, the low crystallinity (inferred from the relatively low intensity of the XRD peaks) is observed in Sm$_{0.8}$Sr$_{0.2}$NiO$_2$, which is a common feature of soft chemical reduction method \cite{soft}. And there are also a few broad peaks which can be attributed to the low crystallinity and the buckling of NiO$_2$ planes. The cell parameters obtained from Rietveld refinement profile are \emph{a} = \emph{b} = 3.8885(3) $\mathring{A}$ and \emph{c} = 3.2572(0) $\mathring{A}$. The reliability factors and the goodness of fit are R$_{wp}$ = 3.46$\%$, R$_p$ = 2.75$\%$ and GOF = 1.09. The lattice constants are smaller than those of bulk Nd$_{0.8}$Sr$_{0.2}$NiO$_2$, which is reasonable due to the smaller ionic radius of Sm$^{3+}$ than that of Nd$^{3+}$. Fig.~\ref{fig5}(b) shows the SEM image of Sm$_{0.8}$Sr$_{0.2}$NiO$_2$, which contains the main Sm$_{1-y}$Sr$_{y}$Ni$_{1-\delta}$O$_2$ grains and a small amount of Ni impurities from the EDS analyses. Using the same way as that for Sm$_{1-x}$Sr$_x$NiO$_3$ (\emph{x} = 0, 0.2) to analyze the EDS data, we display the Ni occupancy of 18 spots in Fig.~\ref{fig5}(c). The mean value of Ni occupancy is 0.96(6), which is close to that in bulk Nd$_{0.8}$Sr$_{0.2}$NiO$_2$ \cite{NdSrNiO bulk}. Fig.~\ref{fig5}(d) shows the compositional distribution of Sr content with Gaussian fitting for Sm$_{0.8}$Sr$_{0.2}$NiO$_2$. The Sr content is randomly distributed in Sm$_{0.8}$Sr$_{0.2}$NiO$_2$ and the average value obtained from the fitting is 0.165$\pm$0.040. Once again, there is no direct relationship between the Ni occupancy and Sr content.

\subsubsection{Magnetic and electrical transport properties}

\begin{figure}
  \centering
  \includegraphics[width=4in]{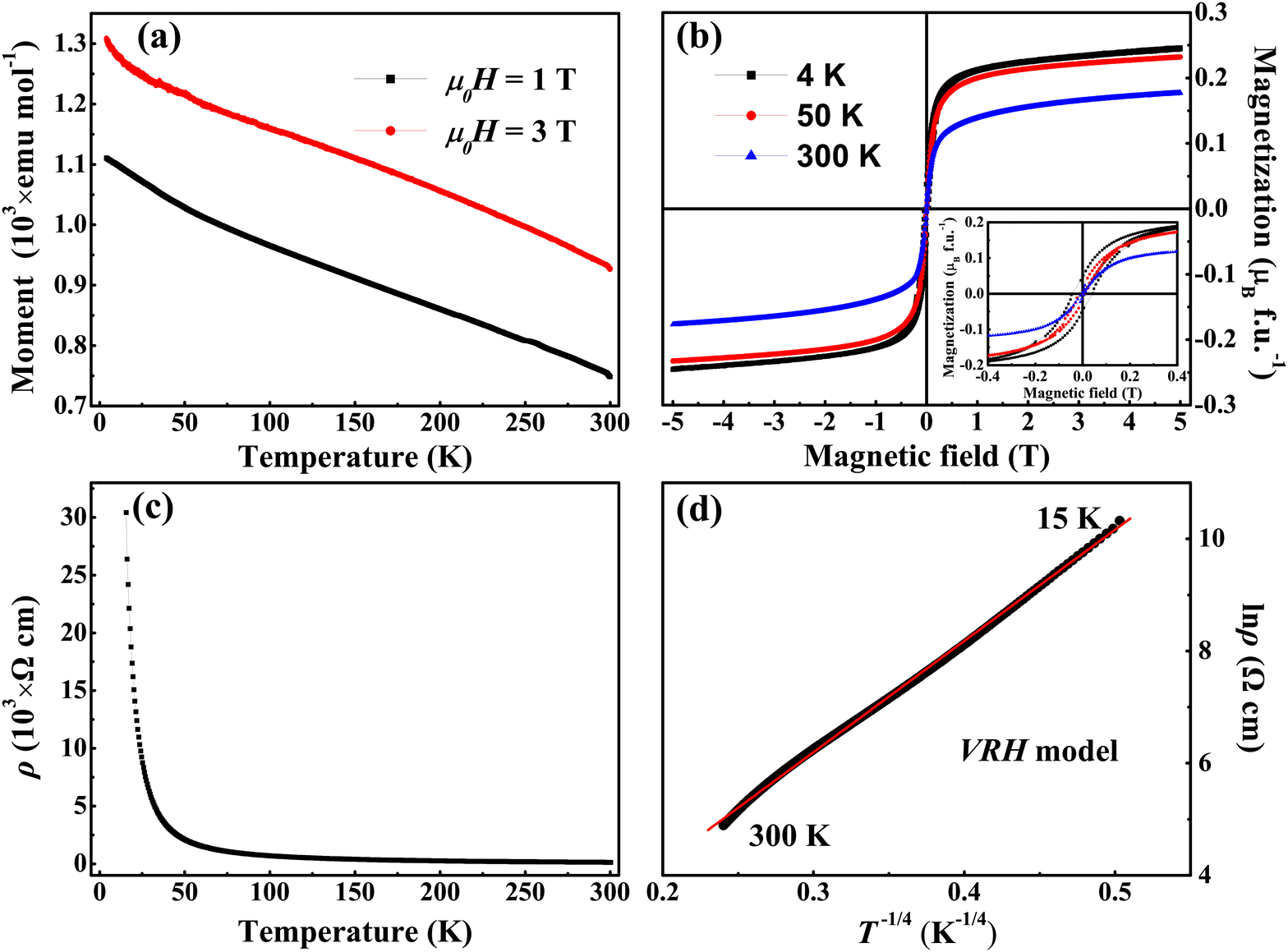}
  \caption{(a) Temperature dependences of magnetic moment measured in ZFC mode at 1 T and 3 T for Sm$_{0.8}$Sr$_{0.2}$NiO$_2$. (b) Magnetization hysteresis loops measured at 4 K, 50 K and 300 K. The inset of (b) shows the zoom-in view of the loops at low fields. (c) Temperature dependence of resistivity for Sm$_{0.8}$Sr$_{0.2}$NiO$_2$. (d) The curve of ln$\rho$ versus \emph{T}$^{-1/4}$ (VRH model).} \label{fig6}
\end{figure}

Figure~\ref{fig6}(a) shows the temperature dependences of magnetic moment measured in ZFC mode under the fields of 1 T and 3 T. The magnetic moment increases almost linearly with the decrease of temperature at 1 T and 3 T. And the two curves are roughly parallel to each other. This is different from the behavior of bulk Nd$_{0.8}$Sr$_{0.2}$NiO$_2$ \cite{NdSrNiO bulk}, in which a paramagnetic Curie-Weiss tail was observed at low temperatures. The magnetization hysteresis loops, shown in Fig.~\ref{fig6}(b), provide further information about the magnetic behavior. The \emph{M}-\emph{H} curves show clear S-shaped hysteresis loops, indicating a weak ferromagnetic component in the paramagnetic background. As in bulk Nd$_{0.8}$Sr$_{0.2}$NiO$_2$ \cite{NdSrNiO bulk}, the ferromagnetic and paramagnetic component can be attributed to the segregated Ni impurities and the main Sm$_{0.8}$Sr$_{0.2}$NiO$_2$ phase, respectively. The different behaviors of magnetic properties between Sm$_{0.8}$Sr$_{0.2}$NiO$_2$ and Nd$_{0.8}$Sr$_{0.2}$NiO$_2$ may be attributed to the different proportion of Ni impurities in the two samples or the intrinsic properties originated from Sm$_{0.8}$Sr$_{0.2}$NiO$_2$. In current situation, the ferromagnetic moment contributed by the Ni impurity seems to be close to that of the bulk Nd$_{0.8}$Sr$_{0.2}$NiO$_2$, while the low temperature Curie-Weiss term seems to be much smaller, suggesting a smaller paramagnetic contribution in Sm$_{0.8}$Sr$_{0.2}$NiO$_2$. Further study is needed to clarify the origin of the magnetic behaviors in Sm$_{0.8}$Sr$_{0.2}$NiO$_2$.

Figure~\ref{fig6}(c) shows the temperature dependence of electrical resistivity for Sm$_{0.8}$Sr$_{0.2}$NiO$_2$. Clearly the sample Sm$_{0.8}$Sr$_{0.2}$NiO$_2$ exhibits an insulating behavior. We have also tried to fit the data using different models in order to reveal the nature of its insulating behavior and find that the $\rho$(\emph{T}) curve can only be well fitted in the whole measured temperature range by the 3D VRH model (shown in Fig.~\ref{fig6}(d)), which has also been observed in bulk Nd$_{0.8}$Sr$_{0.2}$NiO$_2$ \cite{NdSrNiO bulk}, NdNiO$_3$ thin films \cite{VRH2} and Nd$_{1.67}$Sr$_{0.33}$NiO$_4$ polycrystals \cite{VRH3}. The VRH conduction indicates a strong localization behavior and usually occurs in disordered or correlated systems \cite{VRH4}. The VRH conduction observed here suggests the existence of disorder and/or strong correlation effect in Sm$_{0.8}$Sr$_{0.2}$NiO$_2$. The insulating behavior may get reasons from grain boundary scattering and disorders in the synthesized samples. There is still no sign of superconductivity in Sm$_{0.8}$Sr$_{0.2}$NiO$_2$.

\section{Conclusions}

In summary, we have successfully synthesized the perovskite Sm$_{1-x}$Sr$_x$NiO$_3$ (\emph{x} = 0, 0.2) and infinite-layer Sm$_{0.8}$Sr$_{0.2}$NiO$_2$ polycrystalline samples. From XRD and EDS analyses, the Ni occupancy in all the three samples deviates slightly from the nominal value and the Sr content is unevenly distributed in Sm$_{0.8}$Sr$_{0.2}$NiO$_3$ and Sm$_{0.8}$Sr$_{0.2}$NiO$_2$. The sample SmNiO$_3$ undergoes an antiferromagnetic transition at about 224 K on an insulating background. However after Sr doping, Sm$_{0.8}$Sr$_{0.2}$NiO$_3$ becomes a metal and a possible spin-glass state is observed below 12 K from the magnetization measurements. The Sm$_{0.8}$Sr$_{0.2}$NiO$_2$ sample exhibits an insulating behavior and superconductivity is still absent in the present compound. These results provide some useful information of the Sm derivative nickelates, and also clues for the realization of superconductivity in bulk nickelates.

\section*{Acknowledgments}
This work was supported by the National Key R\&D Program of China (Grant No. 2016YFA0300401 and 2016YFA0401704), National Natural Science Foundation of China (Grant No. A0402/11534005 and E0209/52072170), and the Strategic Priority Research Program of Chinese Academy of Sciences (Grant No. XDB25000000).

\end{document}